\documentstyle[epsf,rotate]{mn}

\title[giant swallowing model for V838~Mon]
{A model of an expanding giant that swallowed planets for the eruption of
V838~Monocerotis}
\author[Retter \& Marom]{
\parbox[t]{\textwidth}{
\vspace{-1.0cm}
Alon Retter\thanks{Email: retter@physics.usyd.edu.au} and
Ariel Marom\thanks{Email: amarom@physics.usyd.edu.au}
}
\vspace*{6pt} \\ 
School of Physics, University of Sydney, 2006, Australia\\
\vspace*{-0.5cm}}

\date{Accepted ???. Received ???; in original form ???}

\begin{document}

\maketitle

\begin{abstract}

In early 2002 V838~Monocerotis had an extraordinary outburst whose nature 
is still unclear. The optical light curve showed at least three peaks and 
imaging revealed a light echo around the object -- evidence for a dust 
shell which was emitted several thousand years ago and now reflecting light 
from the eruption. Spectral analysis suggests that the object was relatively 
cold throughout the event, which was characterized by an expansion to 
extremely large radii. We show that the three peaks in the light curve have 
a similar shape and thus it seems likely that a certain phenomenon was three 
times repeated. Our suggestion that the outburst was caused by the expansion 
of a red giant, followed by the successive swallowing of three relatively 
massive planets in close orbits, supplies a simple explanation to all 
observed peculiarities of this intriguing object. 

\end{abstract}


\section{Introduction}

% \subsection{The AM Canum Venaticorum Systems}

The outbust of V838~Mon was discovered in 2002 January (Brown 2002). The 
object was about 6 mag brighter than its presumable quiescent 
brightness level (Munari et el. 2002a). After a short phase of slow 
decline it had a second episode of fast brightening by nearly the same 
factor. V838~Mon gradually faded again in February, but in March a third 
peak was observed. Figs.~1 \& 2 display the visual light curve of the 
outburst compiled from various sources. To our knowledge such a light 
curve has never been observed before. 

% Evans et al. (2003) used IRAS and 2MASS data to conclude that the 
% temperature of the progenitor of V838~Mon was 7300~K. 

Spectra of V838~Mon taken by several groups (Munari et el. 2002a; 
Goranskii et el. 2002; Banerjee \& Ashok 2002; Wisniewski et al. 2003) 
suggested a very cool and extended photosphere throughout the eruption. 
V838~Mon became redder during the outburst and the following decline. 
The spectrum near the peak of the outbust was fitted with a K5 giant 
while several months later an M10 giant can be concluded (Munari et el. 
2002b; Munari \& Desidera 2002). Later infrared observations in 2002 
October led to the suggestion that this star is the first L-supergiant 
(Evans et al. 2003). This evolution in time corresponds to a gradual 
decrease in the effective temperature from $\sim$6000~K to less than 
$\sim$2300~K (Munari et el. 2002b; Kimeswenger et al. 2002; Evans et al. 
2003). 

The light echo observed around V838~Mon is an indication of high mass 
loss, which occurred several thousand years ago (Munari et el. 2002a; 
Bond et el. 2003). The distance to the object is still uncertain. 
Initial estimates varied from 0.64~kpc (Kimeswenger et al. 2002) to 
$\sim$10~kpc (Munari \& Desidera 2002). Later and more detailed 
analysis suggests that the distance is at least 5-8~kpc (Bond et al. 
2003; Crause et al. 2003a; Tylenda 2003). Assuming a distance around 
8~kpc, the estimated maximum expansion radius of V838~Mon (Soker \& 
Tylenda 2003) is $\sim$3200~$R_{\odot}$ ($\sim$15~AU). 

% It was estimated that the radius of V838~Mon was increased by a factor 
% of $\sim$10 from after the first brightening in January till April 
% (Munari et el. 2002a; Soker \& Tylenda 2003). 

Evans et al. (2003) used IRAS and 2MASS data to conclude that the 
temperature of the progenitor of V838~Mon was $\sim$7300~K. Its 
luminosity at 8~kpc is then $\sim$160~L$_{\odot}$. Combining these two 
parameters, we find a stellar radius of $\sim$8~$R_{\odot}$ for the 
progenitor of V838~Mon, which would suit a star on the red giant 
branch (RGB).

% according to the scaled equation of Evans et al. (2003). 

% at least 250~$R_{\odot}$ assuming the minimal distance, but it is likely 
% that the distance is around 6~kpc (Bond et al. 2003) corresponding to a 
% maximum expansion radius of $\sim$2400~$R_{\odot}$.

\begin{figure*}

\centerline{\epsfxsize=4.5in\epsfbox{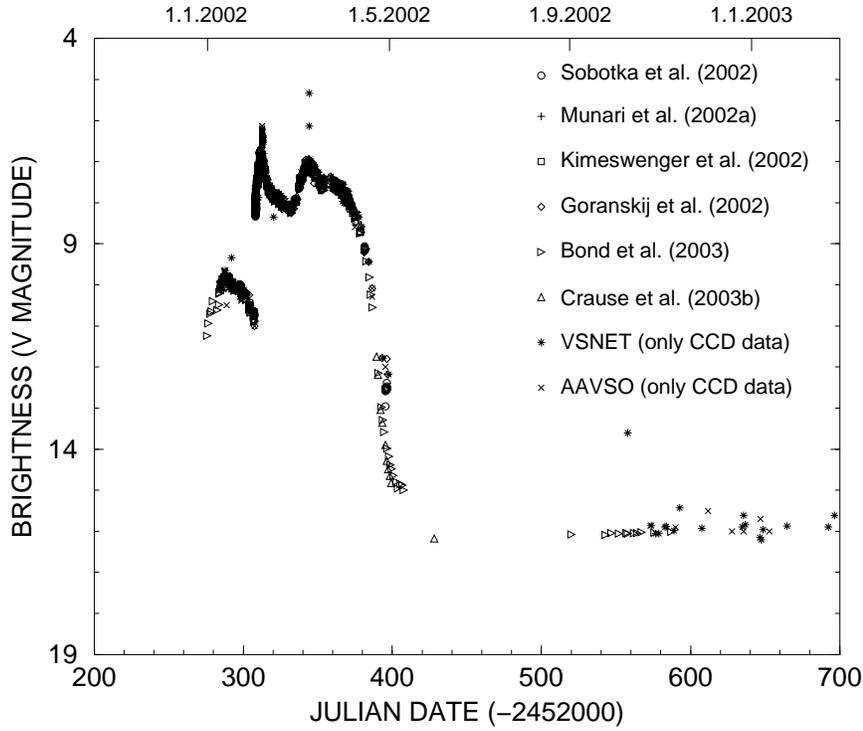}}
%\centerline{\epsfxsize=3.0in\epsfbox{fig1.eps}}

\caption{The optical light curve of V838~Mon in the visual band compiled
from various sources. The light curve comprises more than one thousand CCD 
measurements. The outburst occurred during the first four months of 2002,
after which the object settled to V$\sim$16.} 

\end{figure*}

\begin{figure*}

\centerline{\epsfxsize=4.5in\epsfbox{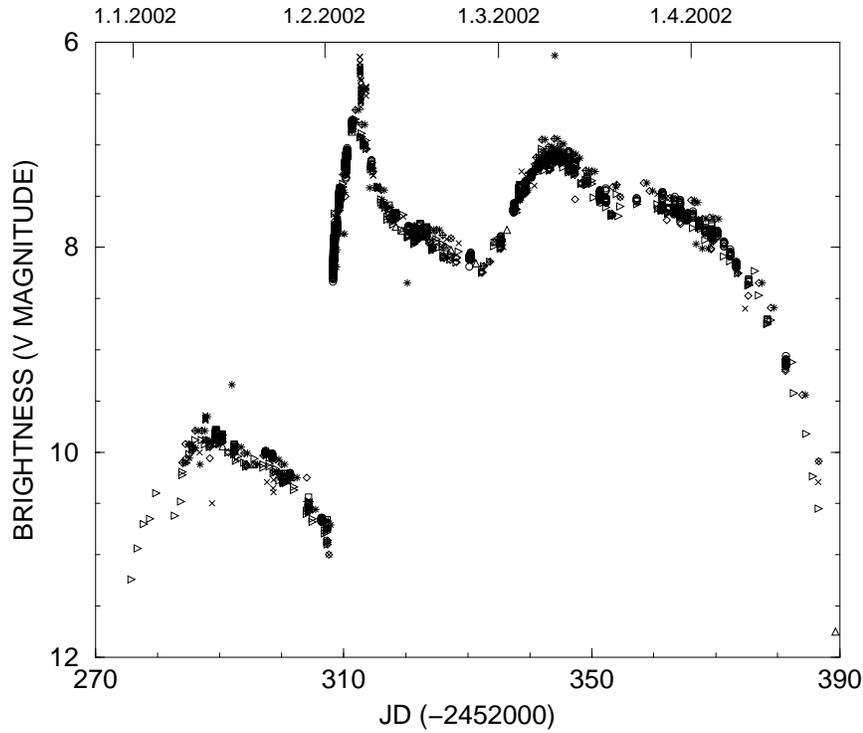}}
%\centerline{\epsfxsize=3.0in\epsfbox{fig2.eps}}

\caption{The structure of the peaks (same as Fig.~1 zoomed into the 
event of the outburst). In January (JD$\sim$2452275-305), February 
(JD$\sim$2452310-335) and March (JD$\sim$2452340-370) the light curve 
showed three peaks with a complicated structure, which can be described 
as composed of an early bright peak accompanied by a secondary weak 
bump about 10-20 days later.} 

\end{figure*}

The overall observed behaviour of V838~Mon is very different from nova 
outbursts and hard to be understood by any of the known astronomical 
events. A bold suggestion that this outburst was caused by the release 
of gravitational energy during a binary merger event has been put 
forth (Soker \& Tylenda 2003). According to this scenario, the first 
brightening was caused by the merger of the outer layers of the two 
stars. Then, after about a month, the cores collided generating more 
energy and causing the second burst. The third peak was ascribed to 
changes in luminosity as the system settled into equilibrium. The 
analysis shows that the gravitational energy released by an accreted 
mass of $\sim$0.05~$M_{\odot}$ is sufficient to inflate the envelope 
of a 1.5~$M_{\odot}$ main-sequence star to $\sim$500~$R_{\odot}$. 

% \section{The complicated light curve of  V838~Mon}

A careful inspection of the light curve of V838~Mon (Figs. 1 \& 2) shows 
that the three peaks have a similar structure, namely each maximum is 
followed by a decline and a very weak secondary peak. In our opinion this 
characteristic of the light curve is inconsistent with the binary merger 
model mentioned above. The shape of the light curve prompts us to argue 
that V838~Mon had three events of similar nature, but probably of different 
strengths. The obvious candidate for such behaviour is the swallowing of 
massive planets in close orbits around a parent star.

During the past decade radial velocity studies of nearby stars have been 
very successful in finding planets. So far more than one hundred 
candidates for extrasolar planets have been discovered. Massive planets 
of Jupiter's mass (M$_J$) or larger have been seen to be common around 
many stars at very close orbits (Livio \& Soker 2002; Schneider 2003). 

% (http://cfa-www.harvard.edu/planets/cat1.html; Livio \& Soker 2002).

The fate of a planet in close orbit to an expanding RGB or asymptotic 
giant branch (AGB) star has been investigated by several authors (Livio 
\& Soker 1984; Harpaz \& Soker 1994; Soker 1996a,b; 1998; Siess \& Livio 
1999a,b). It is generally accepted that some of the inner planets 
surrounding a star should be engulfed and swallowed as it expands after 
leaving the main-sequence. It was shown that the inner planets in the 
solar system, up to and including Earth are likely to be eventually 
swallowed by the Sun (Rybicki \& Denis 2001). There were also numerous 
reports of signs that many giant stars have in fact swallowed planets 
during their evolution (Siess \& Livio 1999a,b). It was estimated that 
at least 3.5-9\% of stars should experience enhanced mass loss as a 
result of planets and brown dwarf companions spiralling into their 
envelopes (Livio \& Soker 2002). 

% winds that should cause drag and accretion onto them, both resulting 
% in slowing down their velocities and shrinking their orbits. 

Even before being engulfed, planets can be affected by enhanced solar-like 
winds and tidal interaction which becomes the dominant effect at a close
range. The engulfed planet may be evaporated inside the envelope or overflow
its Roche Lobe while spiralling inwards (Harpaz \& Soker 1994; Soker 1996b, 
1998), but there is a range of masses for which the planet can survive and 
expel the envelope and a range (around 1-10 M$_J$) where it can survive up 
to reaching the stellar core (Livio \& Soker 1984; Soker 1998; Siess \& 
Livio 1999a,b). We note that Sandquist et al. (1998) have conducted 3D 
hydrodynamic simulations of Jupiter-like planets entering the envelope of 
main-sequence stars. They have found the results to be sensitive to the 
stellar mass and to the details of planetary interior models. They 
concluded that in some cases planets less massive than Jupiter can survive 
through to the base of the convective zone (see also Sandquist et al. 2002).

\section{Our model}

We would like to propose then that the multi-stage eruption of V838~Mon 
was the swallowing of three planets in succession. In our scenario, in 
addition to the gravitational energy generated by the process, there may 
also be a rapid release of nuclear energy as `fresh' hydrogen is driven 
into the hydrogen burning shell of a post main-sequence star. The 
material arrives from the outer convective stellar region as a shock 
wave is formed by the infalling supersonic body. This will be 
augmented by material from the incoming planet itself if it is massive 
enough (probably several M$_J$, around 0.01~$M_{\odot}$) to avoid being 
evaporated away before reaching the nuclear burning region. The 
replenishment of the dwindling hydrogen abundance in the burning shell 
can cause a significant increase in the rate of nuclear reactions. 
This should be even more enhanced if the incoming hydrogen manages to 
penetrate deeper into the star where temperatures are higher. 

A rather similar situation has been investigated and the effects of a 
large planet or a brown dwarf companion being swallowed by an AGB or 
RGB star were calculated (Siess \& Livio 1999a,b). It was found that 
the accretion can be accompanied by substantial expansion and subsequent 
ejection of matter, infrared emission, increase in surface metalicity 
(especially Lithium abundance) and a spin-up of the star. The case 
considered was where the planet's mass is deposited into the interior 
of the star at a slow rate and this was represented as an internal 
accretion process. A spherically symmetric stellar evolution code was 
used and accretion rates of only $10^{-5}-10^{-4}$~$M_{\odot}$ 
yr$^{-1}$ were assumed. Even then, completing the rather complicated 
evolution and following the increase in luminosity for the higher mass 
transfer rate used was difficult. The situation we are describing calls 
for matter to enter the hydrogen burning region at higher rates and the 
entire process terminating much sooner than in the above mentioned 
calculations. 

The gravitational energy released by an infalling planet of several
M$_J$ is very large. If we consider a planet with 5~M$_J$ that spirals-in 
into a host star with a core mass of 1~M$_{\odot}$ within a time scale of 
one month, the rate at which gravitational energy is released is:

\begin{equation}
L \sim \frac{G(1 M_{\odot})(5\times0.001 M_{\odot})}{1 R_{\odot}} / 30 d \sim 2\times10^6 L_{\odot}
\end{equation}

This is much larger than the luminosity emitted in the first peak, which 
is $\sim 3\times10^4 L_{\odot}$ for a distance of 8~kpc (Soker \& Tylenda 
2003), and even bigger than the luminosity of the entire outburst 
($\sim 6\times10^5 L_{\odot}$). On the other hand, the enormous expansion 
of the stellar envelope might require the utilization of most, if not all
the released gravitational energy. In this case, we conclude that most of 
the gravitational energy was probably used to throw away the ejecta and
the luminosity rise was due to nuclear energy. 

While the scenario we are promoting could occur in any of the post 
main-sequence stages from the RGB to the AGB phase, it appears to most 
likely happen at the RGB stage. At this phase, stars with masses up to 
2~$M_{\odot}$ (i.e. most stars) will expand to almost the same radius 
they will reach later during the AGB phase and the thermal pulses (Siess 
\& Livio 1999a,b; Rybicki \& Denis 2001). Thus, any planet that is 
eventually engulfed by its host star is likely to be accreted at this 
earlier phase. We note that as stated in Section~1, the estimated radius
of V838~Mon prior to the eruption is suitable for the RGB phase.

It should be emphasised that the initial engulfment could be as a result 
of the very slow expansion of the giant star and then a rapid eruption 
consequently taking place, or a stellar flare (shell hydrogen burning 
tends to be unstable) that is greatly enhanced by the swallowed planet 
(or brown dwarf). The induced eruption causes additional planets to be 
engulfed and swallowed. The material that was seen to have been previously 
ejected from the star (the light echo, Section~1), could well have been a 
result of an even inner planet that had been accreted at an earlier stage. 
That event should not have been strong and extensive enough to totally 
engulf the planets that were swallowed during the present outburst, but 
may have started their spiralling-in process a long time ago. 

The free fall time for a planet with an initial separation, $a$, into a 
star with a mass $M$ is given by:

\begin{equation}
t_{ff}=7.5 d (\frac{M} {2{M_{\odot}} } )^{-1/2} (\frac{a}{0.3 AU})^{3/2}
\end{equation}

%t$_{ff}$=7.5 d $(\frac{M} {2{M_{\odot}} } )^{-1/2} (\frac{a}{0.3 AU})^{3/2}$

Assuming that V838~Mon has a mass in the range of 1-3 M$_{\odot}$ and 
that the actual spiralling-in time is within an order of magnitude of 
this obvious lower limit, the timing of the second and third peaks 
relative to the initial outburst would constrain all three planets to 
be within about 0.5 AU from the star just prior to the outburst (which 
is obtained assuming the actual falling-in time is about 4 times the 
free-fall time and taking into account that the last peak was about 
two months after the initial burst). Returning once again to the 
estimated value of the progenitor's radius, being within the range 
expected for a star in the RGB (Section~1), we find that this value 
is also consistent with what would be expected for a strong tidal 
interaction with planets in the range described by Eq.~2 (Soker, 
private communication) and is thus in line with our model. 

% Assuming V838~Mon has a mass in the range of 1-3 M$_{\odot}$ (though it 
% could be even more massive judging from its enormous expansion), a simple 
% estimate of the free fall times would constrain all three planets to be 
% within around one AU from the star just prior to the outburst.

It is believed that Lithium is destroyed at early stages in stars as it 
is convected into hot nuclear burning regions. The accretion of planets 
could cause a significant enhancement of this and other elements on the 
surface of stars. While several other mechanisms have been offered to 
explain the Lithium enrichment observed in many evolved giant stars, 
there are problems with these scenarios. Therefore, the presence of this 
element in several giant stars was taken as an indication that these 
objects have swallowed planets in the past (Siess \& Livio 1999a,b). In 
fact, the reported detection of the isotop Li$^6$ in the atmosphere of 
HD~82943, which hosts two planetary systems, was interpreted as evidence 
for accretion of planet/s (Israelian et al. 2001; 2003), but see also 
Reddy et al. (2002). Lithium was actually observed in the spectra of 
V838~Mon during the outburst (Munari et el. 2002a; Goranskii el. 2002). 
These observations, together with the evidence for ejected matter, 
infrared emission excess, the radius of the progenitor and its significant 
expansion during the outburst (Munari et al. 2002a; Banerjee \& Ashok 
2002; Kimeswenger et al. 2002; Bond et al. 2003; Crause et al. 2003b; 
Evans et al. 2003), are consistent with the predictions of the models 
(Siess \& Livio 1999a,b) and thus strongly support the proposed scenario 
for V838~Mon.

\section{Other considerations}
%\section{Is V838~Mon a binary system?}

The second brightening in the light curve of V838~Mon (Figs. 1 \& 2)
was extremely fast and strong compared with the third peak (the 
information on the first maximum is limited as there are no magnitude 
estimates just before the onset of the outburst). The difference 
between the two peaks may be attributed to planets of different masses,
compositions or even to a solid planet (compared with a gaseous one). 

Recent spectra (Munari \& Desidera 2002; Wagner \& Starrfield 2002), 
taken after the decline from outburst, revealed a weak blue continuum that 
may point to the presence of a hot B-type star in addition to the cool 
outbursting object. We note, however, that since the field of V838~Mon 
is quite crowded, the possible companion may actually be a background 
star unrelated with the erupting object. In addition, even if V838~Mon 
is a member in a binary system, it may have no effect on the scenario 
suggested above. Indeed planets in close orbits have been observed in 
binary systems (Dvorak et al. 2003) and investigations have shown that 
they can be stable for a long period of time despite the presence of a 
binary companion (Dvorak et al. 2003; David et al. 2003). 

%\section{V838~Mon as a proto-type of a new class of objects}

The peculiar outburst of V838~Mon may not be unique. A few other stars 
have shown several observational similarities to V838~Mon, and in particular 
similar colour evolution, but the observational data of these objects are 
relatively scarce. The list includes M31RV (Red Variable in M31 in 1988) 
(Rich et al. 1989; Mould et al. 1990; Bryan \& Royer 1992), V4332~Sgr 
(Luminous Variable in Sgr, 1994) (Martini et al. 1999) and maybe also 
CK~Vul (Nova-? Vul 1670) (Shara, Moffat \& Webbink 1985; Kato 2003). It was 
thus suggested that V838~Mon is the proto-type of a new class of objects 
(Munari et al. 2002a,b; Bond et al. 2003), which we propose may all have 
been expanding giants who were observed in the act of swallowing planets.
V838~Mon is presumably an extreme case of this new group, in which three 
relatively massive planets were consumed during the outburst. According to 
our ideas, there should be more examples of expanding giants that swallow 
less and lighter planets thus showing weaker and less spectacular eruptions. 

% Such as example may be V2434 LMC, which showed a peculiar outburst with
% $\sim$3 mag in 2001 (Liller \& Morel 2002).

\section{Acknowledgments}

We are grateful to all professional and amateur astronomers whose data were 
used to plot the figures. In particular we thank P. Sobotka, U. Munari, 
S. Kimeswenger, V. Goranskii, H. Bond and L.A. Crause for sending us their 
photometric data. In this research, we have used, and acknowledge with 
thanks, data from the AAVSO (American Association of Variable Stars 
Observers) and VSNET (Variable Star Network) International Databases, based 
on observations submitted by variable star observers worldwide. We also 
thank K. Wu, T. Mazeh, T. Bedding, L.L. Kiss and especially N. Soker and 
M. Livio for useful comments and discussions. AR is supported by the 
Australian Research Council.

\end{document}